\documentclass[lettersize,journal]{IEEEtran}
\usepackage{amsmath,amsfonts}
\usepackage{algorithmic}
\usepackage{algorithm}
\usepackage{array}
\usepackage{booktabs} 
\usepackage[caption=false,font=normalsize,labelfont=sf,textfont=sf]{subfig}
\usepackage{makecell}
\usepackage{textcomp}
\usepackage{stfloats}
\usepackage{url}
\usepackage{verbatim}
\usepackage{graphicx}
\usepackage{color,soul}
\usepackage{cite}
\usepackage{amsmath}
\begin{document}

\title{Toward using Speech to Sense Student Emotion in Remote Learning Environments}

\author{Sargam Vyas$^*$, Bogdan Vlasenko$^*$, Andr\'{e} Mayoraz, Egon Werlen, Per Bergamin, Mathew Magimai.-Doss
\thanks{$^*$S. Vyas and B. Vlasenko contributed equally to this work.}%
\thanks{S. Vyas,  A. Mayoraz and M. Magimai-Doss are with Idiap Research Institute, Martigny, Switzerland.  e-mail: mathew@idiap.ch}
\thanks{B. Vlasenko is with Idiap Research Institute, Martigny, Switzerland; Radiology Department, Lausanne University Hospital and University of Lausanne, Switzerland;
The Sense Innovation and Research Center, Lausanne and Sion, Switzerland}
\thanks{E. Werlen and P. Bergamin are with FFHS: Swiss Distance University of Applied Sciences, Brig, Switzerland}}




\maketitle

\begin{abstract}

With advancements in multimodal communication technologies, remote learning environments such as, distance universities are increasing. Remote learning typically happens asynchronously. As a consequence, unlike face-to-face in-person classroom teaching, this lacks availability of sufficient emotional cues for making learning a pleasant experience. Motivated by advances made in the paralinguistic speech processing community on emotion prediction, in this paper we explore use of speech for sensing students' emotions by building upon speech-based self-control tasks developed 
to aid effective remote learning. More precisely, we investigate: (a) whether speech acquired through self-control tasks exhibit perceptible variation along valence, arousal, and dominance dimensions? and (b) whether those dimensional emotion variations can be automatically predicted? We address these two research questions by developing a dataset containing spontaneous monologue speech acquired as open responses to self-control tasks and by carrying out subjective listener evaluations and automatic dimensional emotion prediction studies on that dataset. Our investigations indicate that speech-based self-control tasks can be a means to sense student emotion in remote learning environment. This opens potential venues to seamlessly integrate paralinguistic speech processing technologies in the remote learning loop for enhancing learning experiences through instructional design and feedback generation.
\end{abstract}


\section{Introduction}
Emotions play a crucial role in the learning process, as they significantly influence cognitive functions such as attention, memory, and problem-solving~\cite{pekrun2011emotions, mcconnell2022emotions}. This is particularly relevant in asynchronous online learning environments, where the absence of face-to-face interactions makes it challenging to recognize and address learners' emotional states. Computational paralinguistic research has made tremendous progress in the past two decades on emotion prediction~\cite{schuller2018speech}. So, one way to holistically approach this problem is to introduce those advances within the remote learning loop. For example, instructional design has been shown to have a significant impact on emotions experienced during learning. This influence is particularly evident in principles and strategies specifically aimed at designing learning environments to positively affect emotional responses \cite{Linnenbrink2016adaptive, astleitner2000designing}. Automatic emotion analysis could be helpful here. 

There are several approaches in the literature that address the impact of instructional design on emotions. For example, the work of Linnenbrink et al. \cite{Linnenbrink2016adaptive} present five principles aimed at fostering adaptive emotional experiences during learning: a) promoting a sense of competence, b) promoting autonomy, c) using personally relevant and engaging tasks, d) focusing on learning rather than social comparison, and e) enhancing a sense of belonging. Based on this model, it can be generally assumed that self-control tasks with sample answers can support students by providing clear and useful feedback.

Another concept is the FEASP approach \cite{astleitner2000designing}, which describes 20 strategies of instructional design. These strategies aim to minimize fear, envy, and anger while enhancing sympathy and joy. For instance, envy can be reduced through self-assessments, as these avoid comparisons with others. Moreover, evaluations are designed to be consistent and transparent, avoiding privileges. Other similar concepts that incorporate emotional aspects into instructional design include Kansei Engineering \cite{chuah2008kansei}, empathic design \cite{tracey2019empathic}, emotional design \cite{kumar2019exploring}, and the ECOLE (Emotional and Cognitive Aspects of Learning) approach \cite{glaser2005promoting}. However, these approaches often operate at an abstract level. Concrete studies investigating specific task types or contexts, such as voice input or audio recordings in online learning, are rare. And there is a lack of studies investigating open-ended responses.

Having said that, some of the existing works provide initial insights. In a qualitative study on oral presentations in virtual reality environments \cite{hou2022oral}, students exhibited four different emotional responses: increased comfort, frustration with feedback, discomfort with unrealistic simulations, and fear of negative avatar interactions. Another study \cite{Lubis2019positive} investigated how positive emotions can be elicited in a chat-based dialogue system. Positive emotion-promoting responses were integrated into the response corpus of a neural network, resulting in system replies that appeared more natural and contained more positive emotions.

There is also evidence that no instructional design universally provides a pleasant experience for all learners \cite{tomita2022visual}. A study on students’ social, emotional, and motivational responses to changes in instructional design \cite{jarvela2000socio} revealed that reactions strongly depend on the socio-emotional orientation of individual students. Learners’ individual interpretations interact with the features of the respective learning environment.
These findings align with a study that conducted interviews about the use of speech inputs in self-control tasks \cite{werlen2025experiences}. 

These studies highlight that sensing students' emotions automatically in remote learning environment can be useful. However, how to sense that within the learning process/environment is an open research question. Self-control tasks with text inputs or/and speech inputs could be a venue to sense student emotions within the remote learning environment, as emotion can be predicted from text as well as from speech. One of the advantages of speech over text inputs in self-control tasks is that text input may not carry enough emotion information, as the self-control tasks involve knowledge-based open-ended questions. The limited presence of emotional content in impersonal texts was a central topic of a discussion session at a recent conference, with this being illustrated using self‑control tasks as an example \cite{werlen2022squeezing}. Research consistently shows that recognizing emotions in written text is less accurate than in spoken language. Data reported in a 2004 study \cite{chuang2004multi} as well as in a 2025 review \cite{mobbs2025emotion} demonstrate that text‑based emotion recognition yields lower accuracy rates compared to speech‑based emotion recognition. Having said that, a similar question also arises for speech inputs, as the open responses from students are spontaneous monologue speech, which may not be expressive enough for emotion prediction. FFHS Brig, a Swiss distance university, has developed a framework involving self-assessment where students taking a course independently assess their submissions on a learning platform and provide speech responses to open-ended questions. In this paper, we build upon that to investigate:
\begin{enumerate}
\item whether such monologue speech responses for open-ended questions exhibit perceptible emotion variation? We investigate that by developing a dataset by segmenting speech in chunks and carrying out sentiment analysis; subjectively annotating the data using human listeners in terms of dimensional emotion labels, i.e., valence, arousal and dominance (VAD); and analyzing them.
\item whether dimensional emotion information can be reliably predicted? We investigate that by conducting automatic dimensional emotion prediction studies on the human annotated data set resulting from the first investigation.
\end{enumerate}  
To the best of our knowledge, this is one of the first such efforts in the paralinguistic speech processing community. 

The remainder of the paper is organized as follows. Section~\ref{sec:back} provides a background on speech emotion recognition and motivates the present work. Section~\ref{sec:dataset} focuses on data acquisition and dataset development. Section~\ref{sec:Annot} presents the subjective emotion labeling studies. Section~\ref{sec:ExpSet} presents automatic dimensional emotion prediction studies. Finally, we conclude. 

\section{Background and Motivation}
\label{sec:back} 

From paralinguistic speech processing perspective, this paper focuses on speech-based emotion analysis and prediction that relates to addressing three fundamental problems, namely, (a) emotional data collection, (b) emotion labeling and (c) emotion prediction. This section provides relevant background on those three fundamental problems and motivates our work.

\subsection{Emotional data collection}

Emotion prediction is tightly coupled to the application domain based on the intensity of linguistic expressiveness - whether naturalistic, acted, or elicited \cite{batliner2000desperately,schuller2018speech}, the nature of interaction - such as human-to-human dialogues, human-machine interactions, or monologues, and modeling aspects, including temporal resolution~\cite{gunes2013categorical}, which in-turn are application domain-specific. 

From a speech expressivity intensity point of view, the speech emotion prediction has been largely carried out on datasets that can be broadly grouped into acted emotion and spontaneous emotion~\cite{batliner2000desperately}. Acted emotional corpora rely on professional actors to generate emotional speech. To ensure high-quality emotional content, researchers have used emotion perception tests with large groups of listeners to assess data quality~\cite{burkhardt2005database}. These corpora typically focus on subsets of basic emotions, as introduced by Ekman \cite{ekman1992_basic}, and are primarily designed for developing emotion classification methods. On the other hand, an example of an emotionally evocative method is the Wizard-of-Oz paradigm \cite{wahlster2001smartkom}, in which subjects interact with a computer system that is fully or partially controlled by a hidden human pretending to be an autonomous machine \cite{martin2012universal}. In Wizard-of-Oz-based emotional data collection, emotional stimuli are integrated into human-machine interaction systems \cite{wahlster2001smartkom, vlasenko2009heading}. Although Wizard-of-Oz setups produce more natural emotional data, they are significantly less expressive compared to acted emotional speech.

In terms of the nature of interaction, human-to-human conversations provide better conditions for recognizing authentic emotional cues. Multi-speaker conversations and monologues represent two distinct modes of human-to-human interaction, each exhibiting significant variability in the perception and expression of emotions \cite{keltner1999social, burgoon1995interpersonal, mariooryad2013exploring}. The emergence and evolution of emotional cues are highly influenced by interpersonal dynamics and social aspects of interaction. Human-to-human communication creates an environment conducive to both expressing and perceiving emotions, one example being teacher-student interactions \cite{alghamdi2020online}. On the other hand, in speech production, the acoustic characteristics of speakers differ when they engage in conversations compared to when they communicate in monologue mode \cite{mori2011constructing, zhu2006comparing}.

In this paper, our interest lies in sensing student emotion within the remote learning environment or process without using any additional emotion evocative methods.  Toward that, this paper investigates the use of spontaneous monologue speech acquired from learners carrying out self-control tasks as part of their learning.

\subsection{Emotion labeling}
\label{sec:data_ser}

For using speech to sense emotions in a learning environment, we have first to ascertain whether the spontaneous monologue speech collected as part of self-control tasks exhibit emotion level variation. In the literature, emotions are typically modeled as (a) categorical classes, such as anger, contempt, disgust, enjoyment, fear, sadness, surprise, and neutral \cite{ekman1992argument} or (b) as the dimensional representations valence, arousal, and dominance \cite{russell1977evidence}. The dimensional emotion representation is the most suitable for our goal of investigating presence of emotional variation, as this gives a way to subjectively label the data with minimal bias and analyze the variations. More precisely, we do not have a good prior knowledge about what kind of emotions get expressed in the monologue scenario that we are investigating, especially given the fact that we are not using any emotion evocative methods. In the remainder of the section, we provide a short literature survey on dimensional emotion labeling.

The VAM \cite{grimm2008vera} database consists of 947 emotional speech samples from 47 German speakers (11 male / 36 female), which were extracted from 12 broadcasts of the talk show “Vera am Mittag” (in English, “Vera at Noon”). The emotions in this database can be regarded as prompted spontaneous emotions. In the first phase, the shows were transcribed, after that the dialogues were segmented into utterances. A subset of these utterances was used for Valence–Arousal–Dominance annotation by 6 (VAM I) and 17 (VAM II) annotators on a 5-point Self-Assessment Manikin (SAM) scale.

The IEMOCAP \cite{busso2008iemocap} database was developed with acted speech of around 12 hours and 26 minutes where researchers manually segmented sessions into utterances and transcribed text. Each utterance was then labelled by three annotators. The database was annotated in terms of categorical emotion classes as well as in terms of dimensional representation with a 5-point SAM scale. 

The RECOLA is a French-based spontaneous speech database where dimensional emotion has been annotated on a 9-point SAM scale~\cite{ringeval2013recola}.  Instead of using the Feeltrace \cite{cowie2000feeltrace} and Gtrace \cite{cowie2011gtrace} tools, the researchers created their own web-based tool ANNEMO. The data was annotated by giving oral instructions to annotators along with a four-page document explaining the procedure for the annotation task. They measured the inter-rater agreement using mean correlation coefficient and Cronbach's $\alpha$.

The MSP-IMPROV database \cite{MSP-Improv2017Busso} is an English-based audio-visual acted data which has been annotated using crowdsourcing, by building upon experiences gained from development of MSP-PODCAST database~\cite{MSP_Podcast, busso2008iemocap, provost2015umeme, cao2014crema, tarasov2010using}. Crowdsourcing is particularly helpful while dealing with large databases as it reduces the cost of annotators, along with a reduction in time to acquire annotations \cite{snow2008cheap}.  To annotate the data, dialog turns such as uninterrupted utterances or sentences depending on the length was used to manually segment the corpus. For getting more visual information such as facial gestures, a short silence segment was added at the end and beginning of the turn. On the other hand, MSP-PODCAST data was labelled by obtaining segments based on speaker diarization~\cite{MSP_Podcast} . 

Table~\ref{table:sota_database} summarizes these datasets. We can observe that  the affective research community use different number of annotators and different SAM scales, i.e., there is no single gold standard for annotation. By following the best practices for pre-processing and annotation in the literature discussed above, we developed our own methodology for annotating the monologue spontaneous speech.

\begin{table}[!htb]\centering
\renewcommand{\arraystretch}{1.5}
\begin{tabular}{ |c|c|c|c|c|c| } 
 \hline
 {\bf Database } & {\bf date} $\uparrow$ & {\bf SAM} & {\bf {sp\#}} & {\bf duration} & {\bf ann\#}\\ \hline
 VAM \cite{grimm2008vera} & 2008& 5 & 47 & 48m & 17/6 \\ 
 IEMOCAP \cite{busso2008iemocap} & 2008& 5 & 10 & 12h26m & 3  \\ 
 RECOLA \cite{ringeval2013recola} & 2013 & 9 & 46 & 2h50m  &  6\\ 
 MSP-IMPROV \cite{lotfian2017building} & 2017 & 5 & 12 & 9h35m & 5 \\
 MSP-PODCAST \cite{busso2025msp} & 2025& 7 &  $>3'641$ & 409h& 5 \\ \hline
\end{tabular}
\vspace{2mm}
\caption{Overview of emotional corpora with dimensional labels. Abbreviation:  spk\# - number of speakers, ann\# - number of annotators. SAM corresponds to SAM scale }
\vspace{-0.5cm} 
\label{table:sota_database}
\end{table}

\subsection{Dimensional speech emotion prediction}
Due to the sparse number of samples per speaker (VAM: 947 speech samples for 47 speakers), the early dimensional SER techniques utilized knowledge-based acoustic features as input for Support Vector Regressors (SVR) \cite{rahman2012imaps, weninger2013acoustics, sayedelahl2013audio}. With the emergence of larger emotional corpora (IEMOCAP, MSP-IMPROV, MSP-PODCAST) containing dimensional emotional labels, the affective computing community began applying deep neural networks (DNN) for regression on VAD labels.
DNN models with multi-task learning have been used for regression tasks on the MSP-PODCAST database \cite{parthasarathy2017jointly}. Long Short-Term Memory (LSTM) \cite{Recola_LSTM} techniques have been used for dimensional emotion modeling on the RECOLA database. 
With the advancements in deep learning, fine-tuning of pre-trained self-supervised learning models like wav2vec has been proposed for dimensional emotion prediction~\cite{w2v2_MSP}.
On the other hand, it is still common to use SVR regression techniques combined with advanced acoustic feature or neural feature representations for dimensional SER modeling \cite{busso2025msp, vlasenko_icassp2024}. 

In this paper, rather than developing a new machine approach for dimensional emotion prediction, our interest lies in applying existing methods in the context of the second research question.

\section{Self-control task-based data set}
\label{sec:dataset}

In this section, we present the development of a dataset consisting of student speech inputs as open responses within self-control tasks in a remote learning environment.

\subsection{Data collection setup}

The self-control tasks, initially called prompting tasks, were developed at European distance  university, over a period of several years \cite{werlen2014opel, werlen2022self, werlen2025experiences}. 
In the present study, the self-control tasks are used as a standard task format for distance learning. The tasks consist of a sequence of steps: (0) information relevant to the task/question, (1) open question, (2) open answer, (3) assessment of difficulty level or response confidence, (4) sample answer, (5) self-evaluation - comparison of own answer with sample answer, (6) learning reflection on differences between own answer and sample answer (see \cite{werlen2022self}). 

In the original version, the students \textit{wrote} the open answers and the reflection. In the spring semester of 2021, 
five out of eleven self-control tasks with speech input instead of written input on the Moodle learning platform were offered to the students. For that, the university collaborated with a company
that provided the recording and transcription technology. The field for text input was replaced by a field for speech input. To start the speech recording for entering the open response and for entering the learning reflection, the students clicked on a red microphone symbol and entered their response orally. What was spoken was transcribed directly into a text box. After entering the answer, the students clicked again on the microphone symbol that toggled to black and stopped the recording. The technology behind the speech recognition was based on Kaldi\footnote{https://github.com/kaldi-asr/kaldi}. Kaldi is written in C++ and is based on older, reliable speech recognition models (e.g., Hidden Markov Models, Gaussian Mixture Models, Finite State Transducers)~\cite{povey2011kaldi}. The language model was trained by the partner company for Swiss Standard German and supplemented with the vocabulary of the corresponding course. The interface was developed such that after completing the speech input, the students can access the recordings made so far and the corresponding transcriptions via a bar menu, where they could correct, add to and change the transcriptions as required. The students returned to the recording mode via the bar menu. The tool for speech input and transcription was embedded in Moodle using LTI\footnote{https://docs.moodle.org/405/de/LTI{\_}und{\_}Moodle} (Learning Tools Interoperability\textsuperscript{®}). 

\subsection{Data collection and management}
The recordings took place 
during the spring semester of 2021 of an introductory course in project management for business economists (33\%), informatics scientists (35\%), and business informatics specialists (32\%). There were 142 students enrolled in nine classes, 79 of whom gave their consent for their data to be used for research. All students had access to the tasks with speech inputs for open response. In total, 56 out of the 79 students provided speech recordings.

In total, 815 speech recordings with a total duration of 4.7 hours were collected. 
The audio files were stored in \textsc{wav} format with a sampling frequency of $16$ kHz and a precision of $16$ bits. The mean duration of a recording was $21$ seconds (SD = $17$”; min = $0$”’; max = $131$”). All voice recordings were transcribed automatically and partially corrected and changed by the student. In addition to transcripts, the $56$ speaker IDs were pseudonymized. 

Considering the General Data Protection Regulation (GDPR), the student speech recordings were deleted after the final date specified in the data usage agreement had expired.

\subsection{Data preprocessing and selection}
Annotation of the whole 4.7 hours data can lead to cognitive overload on human annotators, which in turn can lead to annotation issues. As presented earlier, the speech is transcribed automatically and then corrected by the students.  So, text-based sentiment analysis could be used for reducing the data size.

We first segmented the recordings. Considering that emotions are short-term states, we chose to split sentences based on long silence segments. Hence, during the initial stages of data pre-processing, we employed the Montreal Forced Aligner (MFA) \cite{mcauliffe2017montreal} for the force-alignment of our standard German audio dataset. The MFA was selected due to the absence of aligners specifically tailored for standard German. 
Utilizing a German dictionary, MFA employs a triphone acoustic model developed on the Kaldi framework. Comparative studies have demonstrated MFA's superior performance over Prosodylab-Aligner (PLA) and FAVE, which rely on monophone models for force alignment~\cite{mcauliffe2017montreal, gonzalez2020comparing}. The MFA generates an output file in textgrid format for each corresponding .wav file. These textgrid files encapsulate word-to-phoneme alignments along with their respective time intervals. Subsequent to word-level alignment, we employed periods ('.') as delimiters to concatenate words into phrases. This data was then organized by pseudonymized speaker ID, facilitating later stages of phrase annotation. Additionally, timestamps were generated for each phrase for the same purpose. Based on the obtained word/phoneme alignment, we parsed long monologue speech recordings into short, semantically completed chunks. Long silences were used as indicators for our chunking approach.

\begin{figure}[!htb]
    \centering{
    \includegraphics[width=0.5\textwidth]{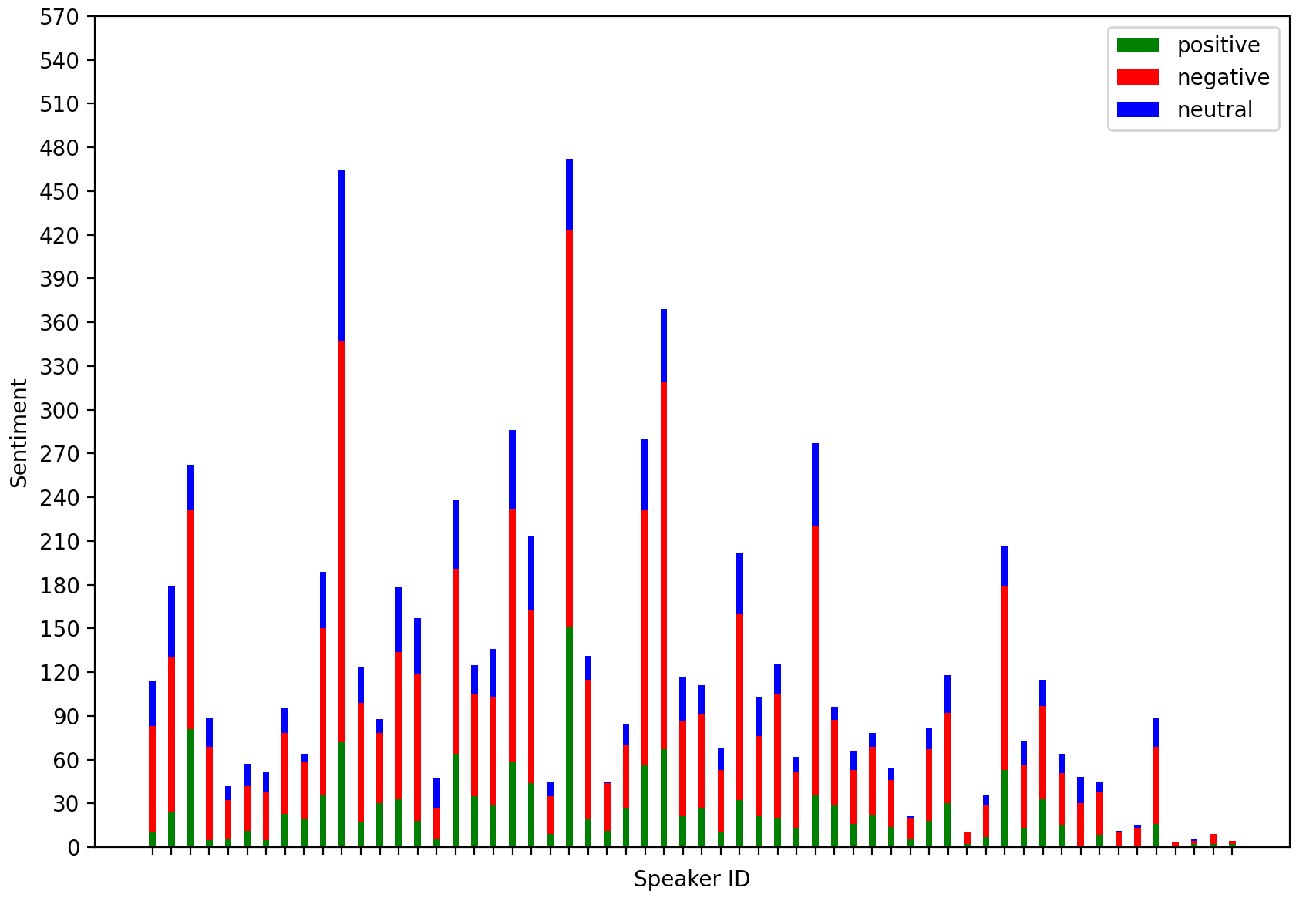}
    }\caption{Distribution of sentiment labels: positive, negative and neutral for $7$k segments. Speaker-wise combined histogram. } 
    \label{fig:sentiment}
\end{figure}

We then conducted sentiment analysis on the extracted 7k transcripts for selected semantically completed chunks. For that, we utilized a German sentiment analysis model based on the BERT architecture~\cite{devlin2019bert}. The model, detailed in the work by \cite{guhr2020training}, was originally trained on a corpus of 1.834 million German-language samples. The distribution of {\it negative}, {\it positive}, and {\it neutral} sentiments for each speaker in our corpora can be found in Figure \ref{fig:sentiment}. Through this step, we could achieve a balanced representation of emotional variability across different speakers in our annotation selection process.

Finally, we chose 1'132 speech wave chunks, aiming to include phrases evenly from all speakers, as the distribution of phrases was not initially balanced. For the timestamp, we sought to create balance by avoiding segments that were too short (where emotions might not be clear) or too long (where emotions could get mixed up). For sentiments, we used a 4:4:2 ratio, with four positive, four negative, and two neutral sentiment predictions. 

We refer to the resulting 56 speaker dataset containing 1 hour 21 minutes of speech as SPoken Online Tasks - Emotions Database \textsc{SPOT-ED}.

\section{Subjective emotion assessment}
\label{sec:Annot}

To ascertain whether there is a perceptible emotion variation along the three dimensions valence, arousal and dominance, we recruited six native listeners. Of the six raters, four were male and two were female. At the time of annotation, their ages ranged from 33 to 57 years (M = 44, SD = 9.5). Four raters were trained psychologists, one was a linguist, and one had a background in education.

\subsection{Annotator training using VAM corpus}
As we observed in Section~\ref{sec:data_ser} no standard techniques exist for training annotators to label spontaneous emotions in the VAD dimensionality. So, we provided a detailed description of emotional dimensionalities before conducting the A/B test using data from VAM corpus, as German dataset which has been labeled along VAD dimensions (see Section~\ref{sec:data_ser}). This approach, combined with the A/B test, allowed annotators to become familiar with the SAM images for each emotional dimension, while preparing them for reliable annotation. 

\begin{figure}[!htb]
    \centering{
    \includegraphics[width=0.5\textwidth]{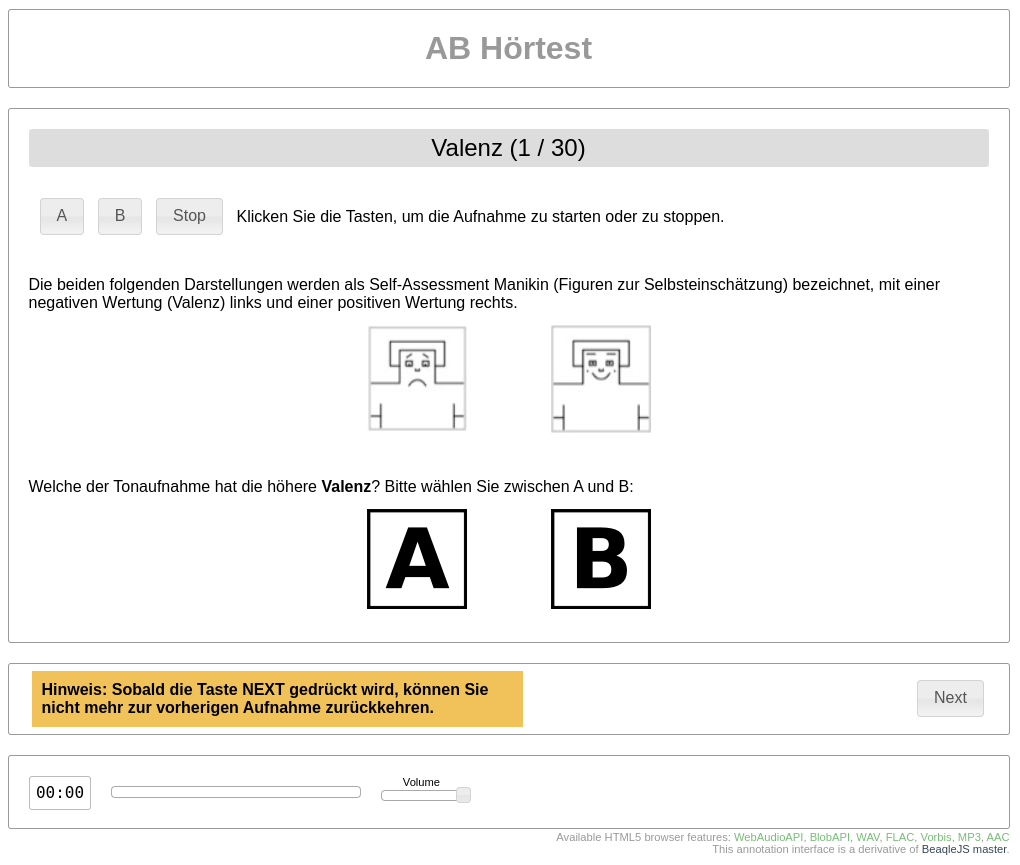}
    }\caption{The AB test for the evaluation of emotion annotation skills.}
    \label{fig:AB_test_annotators}
\end{figure}

To evaluate the emotional speech annotation skills of annotators, we adopted the A/B test proposed in BeaqleJS framework, as shown in Figure \ref{fig:AB_test_annotators}. We selected twelve pairs of emotional speech samples from the VAM database for arousal and valence emotional dimensions, while for the dominance dimension, six pairs of speech samples were used. During the selection of emotional speech sample pairs, we focused on emotional instances with the lowest/highest possible aggregated labels to ensure a diverse range of emotions. 

Table \ref{table:AB_test_annotators} displays the percentage of correct assessments by annotators. As one could see from the table, the percentage of correct assignments can be different for different emotional dimensions for the same annotator.

\begin{table}[!htb]\centering
\renewcommand{\arraystretch}{1.5}
\begin{tabular}{ |c|c|c|c| } 
 \hline
 {\bf ID} & {\bf \textsc{Valence [\%]}} & {\bf \textsc{Arousal [\%]}} & {\bf \textsc{Dominance [\%]}} \\ \hline
 1 & 100 & 92 & 67 \\ 
 2 & 100 & 100 & 100 \\ 
 3 & 42 & 67 & 83 \\ 
 4 & 50 & 83 & 100 \\ 
 5 & 9 & 100 & 100 \\ 
 6 & 100 & 100 & 100 \\ 
 \hline
\end{tabular}
\vspace{2.5mm}
\caption{A/B test results for 6 annotators for Valence, Arousal and Dominance}
\label{table:AB_test_annotators}
\end{table}

\subsection{SPOT-ED annotation}
\label{section:Idiap_annotation}

\begin{figure}[!htb]
    \centering{
    \includegraphics[width=0.5\textwidth]{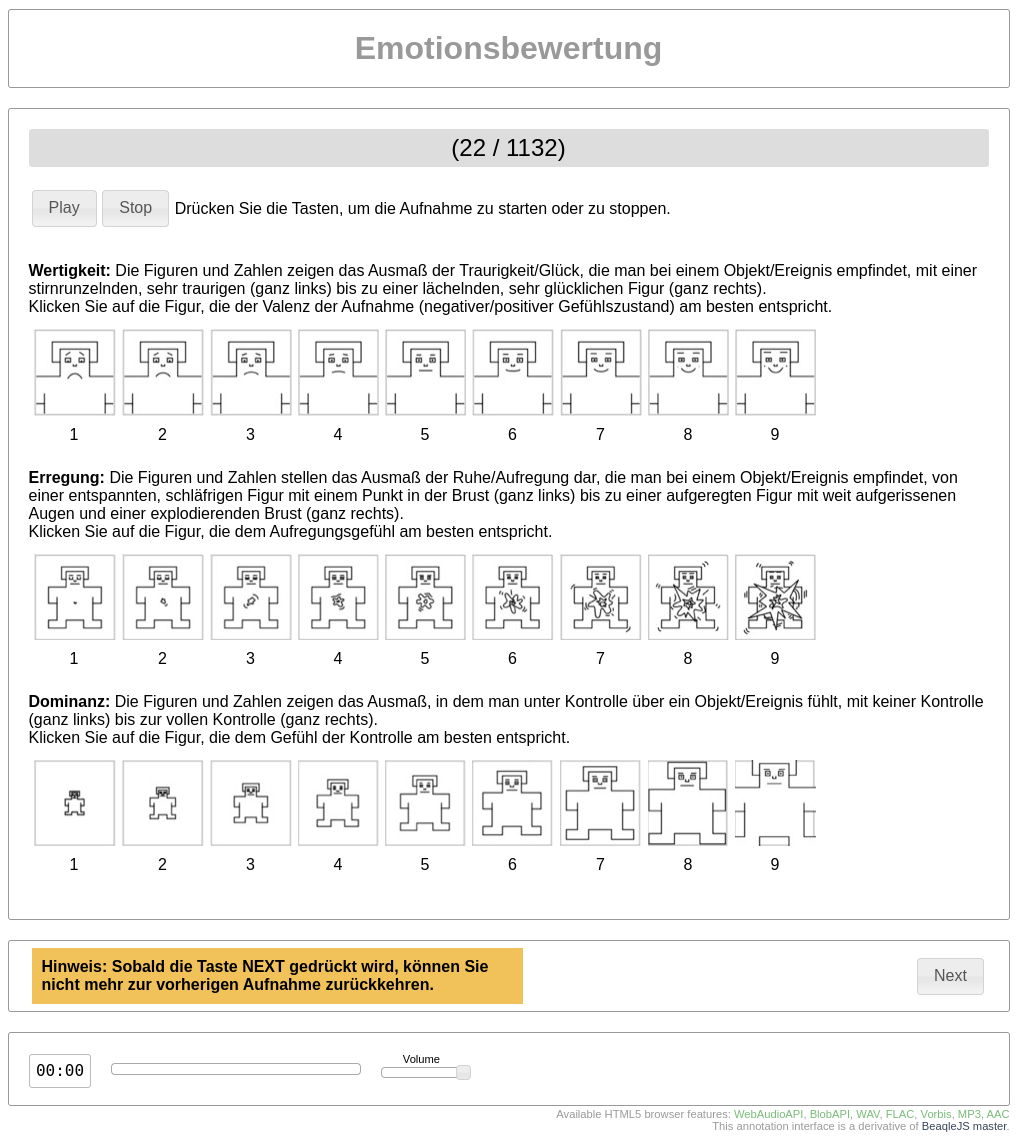}
    }\caption{Self-Assessment Manikin for valence, arousal, dominance ranges with 1 to 9 where 1 being very negative (valence), very calm (arousal) and very weak (dominance). On the other hand, 9 is very positive (Valence), very excited (arousal) and very strong (dominance).}
    \label{fig:SAM}
\end{figure}

For annotating the SPOT-ED data, we used Self-Assessment Manikins (SAM) \cite{BRADLEY1994Measuring, sainz2022gender}. As can be seen from Table~\ref{table:sota_database}, the state-of-the-art databases use different SAM scales depending on their specific goals. We adopted a nine point scale,  with 1 being very negative for valence, very calm for arousal and very weak for dominance and  9 being very positive for valence, very excited for arousal and very strong for dominance. Figure~\ref{fig:SAM} presents the prototype of the SAM scale used for the purpose of annotation.

After familiarizing with the valence, arousal, and dominance dimensions and the SAM framework through the A/B test on the VAM corpus data, the annotators annotated the 1'132 speech chunks of the SPOT-ED dataset. To further avoid cognitive load, the annotators were provided the opportunity to annotate the data iteratively.  After each iteration, the dimensional annotations were stored in a local web-based database. All annotators successfully completed the annotation of the data.

To determine the three dimensional label for each speech wave chunk, we adopted the evaluator weighted estimator (EWE), which has been shown to yield 20\% better evaluation when compared to maximum likelihood estimator~\cite{grimm_kroschel_ewe_2005, grimm2007primitives}. Also, as observed earlier the same annotator can have different levels of annotation capability for different dimensions (see Table~\ref{table:AB_test_annotators}). Briefly, this involved:
\begin{enumerate}
\item Transformation of the three dimensional responses provided by the annotators for each chunk of speech in the range [1~9] to [0~1]\footnote{In~\cite{grimm2007primitives}, the annotator responses were transformed to a range between [-1~+1]. We chose the range [0~1] following recent work on dimensional emotion prediction using state-of-the-art self-supervised learning-based neural network~\cite{w2v2_MSP} and our recent work on VAM and IEMOCAP corpora~\cite{vlasenko_icassp2024}.}, representing negative-to-positive for valence, calm-to-excited for arousal and weak-to-strong for dominance.
\item For each dimension, estimation of an annotator dependent weight, a correlation coefficient that measures the correlation between the annotators responses and the average ratings of all annotators.
\item For each speech wave chunk, taking a weighted sum of the transformed annotators response values based on the dimension specific annotator-dependent weight and normalizing it. This resulted in aggregated EWE value/label for each dimension corresponding to each speech wave chunk.
\end{enumerate}

Following~\cite{grimm2007primitives}, the assessment quality for each speech chunk was determined by estimating the standard deviation between the values based on the response of each annotator and the aggregated EWE value. The inter-annotator agreement then was measured in terms of:
\begin{enumerate}
\item Average of assessment quality, denoted as $\bar{\sigma}$.  A low average indicates that the emotional expression is perceived by all annotators similarly. 
\item Inter-annotator correlation, denoted as $r$, estimated by averaging of the annotator dependent weight (correlation coefficient) across annotators. A high value indicates high inter-annotator agreement.
\end{enumerate}

Table~\ref{table:IRR_VAM_FFHS} presents the inter-annotator correlation $r$ and the average assessment quality $\bar{\sigma}$  for SPOT-ED dataset. To give  an idea of the annotation quality, they are contrasted with VAM I and VAM-II labeling reported in~\cite{grimm2007primitives} with 17 and 6 annotators, respectively. It can be observed that for SPOT-ED for all dimensions $r$ is $\ge 0.6$. In the case of VAM I and VAM II, valence has relatively low agreement compared to arousal and dominance, while the agreement for arousal is high. The reason for that could be that VAM dataset consists of talk shows. Furthermore, we observe that the average assessment quality $\bar{\sigma}$ is low.

 \begin{table}[!htb]
\centering
\renewcommand{\arraystretch}{1.5}
\begin{tabular}{lcccc}
    \toprule
     \textsc{Corpora} && \textsc{Valence} &  \textsc{Arousal} & \textsc{Dominance} \\
     \midrule
     &  \phantom{a} & \multicolumn{3}{c}{inter-annotator correlation $r$ $\uparrow$} \\
     \cmidrule{2-5} 
     \textsc{SPOT-ED} && 0.65 & 0.60 & 0.67 \\
      \textsc{VAM I}~\cite{grimm2007primitives} && 0.49 & 0.78 & 0.68  \\
      \textsc{VAM II}~\cite{grimm2007primitives}  && 0.48 & 0.66 & 0.54  \\
 \bottomrule
     &  \phantom{a} & \multicolumn{3}{c}{Average assessment quality $\bar{\sigma}$ $\downarrow$} \\
     \cmidrule{2-5} 
     \textsc{SPOT-ED} && 0.12 & 0.18 & 0.19 \\
      \textsc{VAM I}~\cite{grimm2007primitives} && 0.30 & 0.38 & 0.33  \\
      \textsc{VAM II}~\cite{grimm2007primitives}  && 0.28 & 0.30 & 0.29 \\
 \bottomrule 
\end{tabular}
\vspace{2.5mm}
\caption{Inter-annotator correlation and average assessment quality for SPOT-ED, VAM I and VAM II.}
\label{table:IRR_VAM_FFHS}
\end{table}

\begin{figure}[!htb]
    \centering{
    \includegraphics[width=0.4\textwidth]{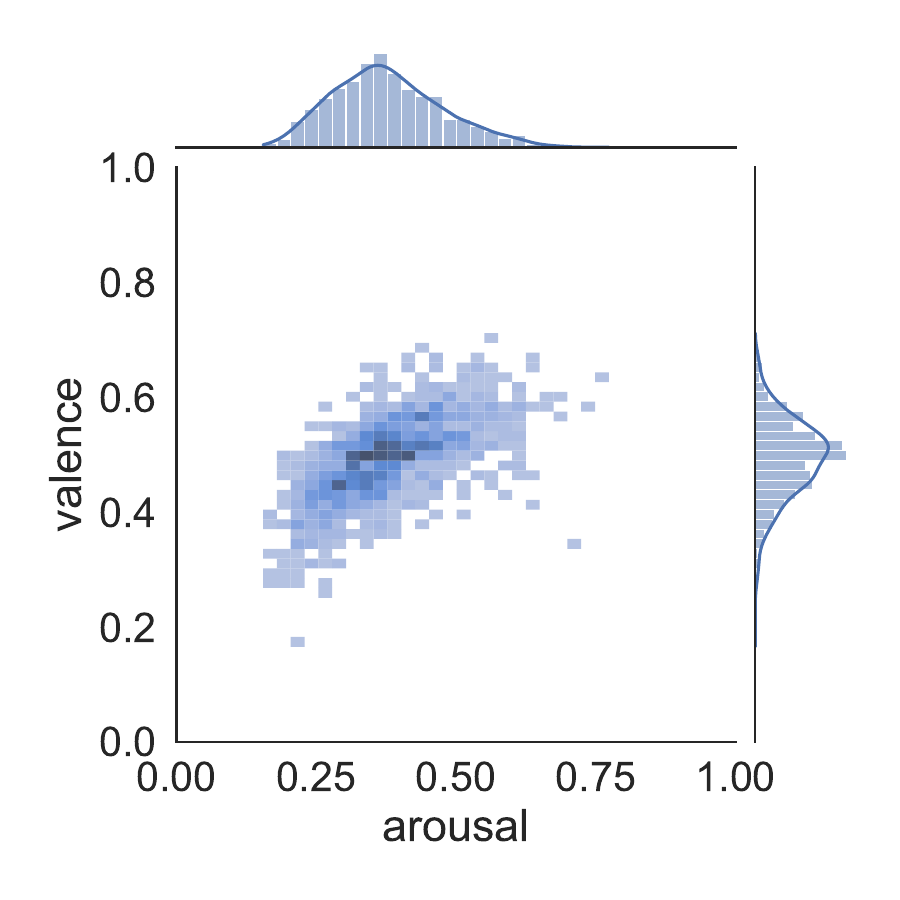}
    }\caption{Distribution of aggregated EWE labels for \textsc{SPOT-ED} database across valence and arousal dimensions.}
    \label{fig:data_dist}
\end{figure}

\begin{figure}[!htb]
    \centering{
    \includegraphics[width=0.4\textwidth]{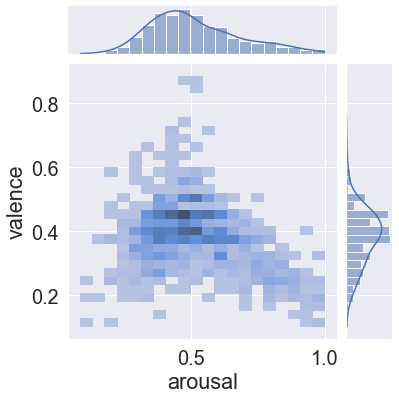}
    }\caption{Distribution of aggregated EWE labels for \textsc{VAM} database across valence and arousal dimensions reported in~\cite{vlasenko_icassp2024}.}
    \label{fig:vam_data_dist}
\end{figure}

Figure~\ref{fig:data_dist} displays the aggregated EWE  labels across valence and arousal dimensions for SPOT-ED. Figure~\ref{fig:vam_data_dist} displays the aggregated EWE labels reported in a recent work of ours on VAM corpus~\cite{vlasenko_icassp2024}. Interestingly, even though SPOT-ED contains only spontaneous monologue speech, it exhibits a wide distribution range similar to VAM, which consists of dialogues from German TV talk show. 

\section{Automatic emotion prediction}
\label{sec:ExpSet}

In the previous section, we found that SPOT-ED exhibits perceptible variability across valence, arousal and dominance dimensions. In this section, given the aggregated EWE emotion labels for the three dimensions, we question: whether we can systematically predict them in an automatic manner? For that, we build upon our recent work on dimensional emotion prediction using knowledge-based features and data-driven pre-trained self-supervised learning based features on VAM and IEMOCAP corpora~\cite{vlasenko_icassp2024}. 

\subsection{Experimental protocol}
We designed two experimental protocols, namely,
\begin{enumerate}
\item Speaker-dependent: in this setup, we trained and tested the dimensional emotion label predictors on the same speaker data. In SPOT-ED, we have 56 unique speakers with approximately 20 speech samples per speaker. We performed 5-fold cross validation study on each speaker data and aggregated the test predictions to compute performance.
\item Speaker-independent: in this setup, we again adopted 5-fold cross validation strategy, where we split the speakers into five non-overlapping groups and carried out Leave-One-Speaker-Group-Out study, i.e., train on four groups and test on the left out group. We aggregated the test predictions to compute performance.
\end{enumerate}
We evaluated the predictors in terms of three performance measures:  Spearman's correlation ($Corr_{spea}$),  Pearson's correlation ($Corr_{pear}$) coefficients, and root mean square error (RMSE). The Spearman correlation coefficient reveals the extent to which the relationship between the model's emotion prediction scores and the subjective human ratings of emotion intensity is monotonic. On the other hand, the Pearson correlation indicates the degree to which the emotion prediction scores are linearly related to human ratings. The RMSE is inversely related to prediction quality; thus, lower RMSE values signify more accurate emotion predictions.

\subsection{Feature representation}

Following our previous work~\cite{vlasenko_icassp2024}, we investigated two types of feature representation (FR), namely,
\begin{enumerate}
\item Knowledge-based: we used Computational Paralinguistics ChallengE (\textsc{ComParE}) \cite{schuller2013interspeech} feature set in Munich open-Source Media Interpretation by a Large feature-space Extraction (\textsc{openSMILE}) \cite{opensmile} toolkit, which applies statistical functionals (e.g., mean, standard deviation, min, max) on frame-level low-level descriptors (LLDs) to output turn-level (utterance-level) $6373$ dimensional representation. This feature representation includes prosody features, spectral features, and cepstral features. 

\item Neural embedding-based: following leader-boards for pre-trained out-of-domain SSL embeddings on emotion recognition sub-challenge for SUPERB challenge \cite{yang2021superb} , we employed (a) \textsc{HuBERT} large\footnote{https://huggingface.co/facebook/hubert-large-ll60k}  (denoted as \textsc{hubert}) \cite{hsu2021hubert_emb}, (b) \textsc{wavLM} large\footnote{https://huggingface.co/microsoft/wavlm-large} (denoted as \textsc{wavlm}) \cite{WavLM}, and (c) \textsc{wav2vec2}  fine-tuned for dimensional emotion prediction\footnote{https://huggingface.co/audeering/wav2vec2-large-robust-12-ft-emotion-msp-dim} on MSP-PODCAST dataset~\cite{wav2vec2_emo} (denoted as \textsc{w2v2-MSP}). 

For each utterance, the utterance level feature representation was obtained by averaging the frame level last transformer layer 1024 dimensional output.
\end{enumerate}

\subsection{Systems}
Similar to our previous work~\cite{vlasenko_icassp2024}, we developed different support vector machine-based regressors for dimensional emotion prediction, namely, (a) stand-alone feature representation-based and (b) feature level combination of knowledge-based representation and neural embedding-based representation. We used  radial basis function-based support vector regression (SVR) with pre-defined non-optimized hyperparameters and with Min Max for feature normalization.

When reporting results, we use numericals for denoting the different systems developed: 1. knowledge-based feature representation, \{2,3,4\} neural embedding feature representation, and $1+x$ for feature level combination where $x \in \{2,3,4\}$.

\begin{table*}[!htb]\centering
\renewcommand{\arraystretch}{1.5}
\small
\begin{tabular}{c c c c c c c c} 
 \hline
&&\multicolumn{3}{c}{\bf \textsc{speaker-independent}}&\multicolumn{3}{c}{\bf \textsc{speaker-dependent}} \\
\cmidrule{3-8} 
{\bf FR} & {\bf Dimension} & {\bf $Corr_{spea}$} $\uparrow$ &  {\bf $Corr_{pear}$} $\uparrow$ & {\bf RMSE} $\downarrow$ & {\bf $Corr_{spea}$} $\uparrow$ &  {\bf $Corr_{pear}$} $\uparrow$ & {\bf RMSE} $\downarrow$ \\ \hline
\multicolumn{8}{c}{\bf \textsc{Standalone feature representations} } \\ \hline
1. \textsc{ComParE}& \textsc{Valence} & 0.429 & 0.439 & 0.064 & 0.489 & 0.511 & 0.061  \\
&\textsc{Arousal} & 0.509 & 0.531 & 0.084 & 0.605 & 0.630 & 0.078 \\
&\textsc{Dominance} & 0.638 & 0.640 & 0.086 & 0.688 & 0.687 & 0.082\\ \hline
2. \textsc{w2v2-MSP} & \textsc{Valence} & 0.525 & 0.551 & \textbf{0.060} & 0.555 & 0.584 & 0.058 \\
& \textsc{Arousal} & 0.587 & 0.611 & 0.080 & 0.621 & 0.657 & 0.076 \\
& \textsc{Dominance} & 0.622 & 0.640 & 0.088 & 0.664 & 0.682 & 0.084 \\ \hline
3. \textsc{HuBERT} & \textsc{Valence} & 0.428 & 0.437 & 0.062& 0.493 & 0.528 & 0.060 \\
& \textsc{Arousal} & 0.503 & 0.532 & 0.085 & 0.607 & 0.642 & 0.079 \\
& \textsc{Dominance} & 0.598 & 0.615 & 0.090 & 0.669 & 0.680 & 0.085 \\ \hline
4. \textsc{WavLM} & \textsc{Valence} & 0.449 & 0.447 & 0.062 & 0.514 & 0.535 & 0.060  \\
& \textsc{Arousal} & 0.510 & 0.552 & 0.084 & 0.599 & 0.642 & 0.078\\
& \textsc{Dominance} & 0.623 & 0.635 & 0.088 & 0.706 & 0.710 & 0.082\\ \hline
\multicolumn{8}{c}{\bf \textsc{Feature level combination} } \\ \hline
1.+2. & \textsc{Valence} & \textbf{0.536} & \textbf{0.562}  & \textbf{0.060} & \textbf{0.585} & \textbf{0.613} & \textbf{0.057} \\
\textsc{ComParE} + &\textsc{Arousal} & \textbf{0.630} & \textbf{0.651} & \textbf{0.076} & \textbf{0.686} & \textbf{0.713} & \textbf{0.072}\\
\textsc{w2v2-MSP} &\textsc{Dominance} & \textbf{0.737} & \textbf{0.744} & \textbf{0.078} & \textbf{0.766} & \textbf{0.767} & \textbf{0.074}\\ \hline
1.+3. & \textsc{Valence} & 0.476  & 0.481 & 0.062 & 0.537  & 0.556 & 0.059 \\
\textsc{ComParE} + & \textsc{Arousal} & 0.554 & 0.578 & 0.082 & 0.643 & 0.667 & 0.076 \\
\textsc{HuBERT} & \textsc{Dominance} & 0.675 & 0.677 & 0.084 & 0.723 & 0.721 & 0.079 \\ \hline
1.+4. & \textsc{Valence} & 0.473 & 0.480 & 0.062 & 0.528 & 0.549 & 0.060 \\
\textsc{ComParE} + & \textsc{Arousal} & 0.548 & 0.573 & 0.082  & 0.64 & 0.664 & 0.076 \\
\textsc{WavLM} & \textsc{Dominance} & 0.678 & 0.679 & 0.084 & 0.726 & 0.723 & 0.079 \\ \hline
\end{tabular}
\vspace{2.5mm}
\caption{Recognition performance for speaker-dependent and -independent experimental setup. Regression models were trained on knowledge-base and data-driven feature representations.}
\label{tab:ffhs_SER}
\end{table*}

\subsection{Results}
\label{sec:Res}

Table \ref{tab:ffhs_SER} presents the performance for the different systems for speaker-independent and speaker-dependent protocols. For the standalone feature representation case, on both setups we observe that \textsc{compare} yields performance comparable to \textsc{hubert} and \textsc{wavlm}, while \textsc{w2v2-MSP} which is dimensional emotion prediction informed tends to yield better performance. When comparing across protocols speaker-dependent systems are typically yielding better values than speaker-independent systems in terms of  $Corr_{spea}$ and $Corr_{pear}$. Having said that, we can also observe that RMSE for the speaker independent in-domain model is as low as 0.060, 0.080, 0.088 for valence, arousal and dominance. Similarly, for the speaker dependent setting, we can observe that RMSE is as low as 0.058, 0.076, 0.084 for valence, arousal and dominance.

For the feature level combination case, we can observe that for all combinations the prediction performance improves over standalone feature based systems. This indicates the complementarity between knowledge-based representation and neural embedding-based representation. We observe that the feature level combination \textsc{ComParE}+\textsc{w2v2-MSP} yields the best system including RMSE as low as 0.060, 0.076, and 0.078 for valence, arousal ,and dominance in the speaker-independent protocol. Furthermore, we also observe that the feature level combination reduces the gap between speaker-dependent prediction and speaker-independent prediction.

\subsection{Analysis}
\label{sec:analysis}

To gain insight into the dimensional emotion acoustic characteristics in the SPOT-ED dataset, similar to~\cite{vlasenko_icassp2024}, for each dimension we calculated a concordance correlation coefficient (CCC) for each feature in the \textsc{compare} feature set after Min-Max normalization and ranked them. Table~\ref{tab:FR_CC_top} presents the top two features with the highest feature-wise \textsc{CCC} for each emotional dimension and contrasts that to the features obtained for VAM in our previous study~\cite{vlasenko_icassp2024}.  Interestingly, we observe that for arousal and dominance the top feature is based on the low level descriptor (LLD) \textbf{audspec{\_}lengthL1norm}. Also in the case of valence, we can see the same trend although different LLDs are ranked high. More precisely, in the case of SPOT-ED \textbf{pcm{\_}fftMag{\_}spectralSlope} is the top ranking feature, whereas in the case of VAM \textbf{mfcc{\_}sma[1]} is the top ranking feature. LLD \textbf{mfcc{\_}sma[1]} is based on the first mel frequency cepstral coefficient, which represents the slope of the smoothed magnitude spectrum. In other words, for both SPOT-ED and VAM the spectral slope-based representation is the top ranking feature for valence prediction.

\begin{table}[!htb]\centering
\renewcommand{\arraystretch}{1.5}
 \begin{tabular}{llc}\toprule
\makecell{\bf \textsc{Dim.}} & {\bf \textsc{Feature representation}}& {\bf \textsc{CCC}} \\ \hline
\multicolumn{3}{c}{\bf \textsc{VAM dataset} \cite{vlasenko_icassp2024}} \\ \hline
{\bf \textsc{Val}} & mfcc{\_}sma[1]{\_}upleveltime75 & 0.199  \\ 
{\bf \textsc{Val}} & mfcc{\_}sma[3]{\_}percentile1.0 & 0.171 \\
{\bf \textsc{Aro}} & audspec{\_}lengthL1norm{\_}sma{\_}quartile3 & 0.725 \\
{\bf \textsc{Aro}} & audspec{\_}lengthL1norm{\_}sma{\_}peakMeanAbs & 0.716 \\
{\bf \textsc{Dom}} & audspec{\_}lengthL1norm{\_}sma{\_}percentile99.0 & 0.659 \\
{\bf \textsc{Dom}} & audspec{\_}lengthL1norm{\_}sma{\_}peakMeanAbs & 0.669 \\ \hline
\multicolumn{3}{c}{\bf \textsc{SPOT-ED dataset}} \\ \hline
{\bf \textsc{Val}} & pcm{\_}fftMag{\_}spectralSlope{\_}sma{\_}peakMeanMeanDist & 0.270 \\
{\bf \textsc{Val}} & pcm{\_}fftMag{\_}spectralRollOff90.0{\_}sma{\_}de{\_}pctlrange0-1 & 0.249 \\
{\bf \textsc{Aro}} & audspec{\_}lengthL1norm{\_}sma{\_}stddev & 0.543 \\
{\bf \textsc{Aro}} & audspec{\_}lengthL1norm{\_}sma{\_}range & 0.541 \\
{\bf \textsc{Dom}} & audspec{\_}lengthL1norm{\_}sma{\_}stddev & 0.460 \\
{\bf \textsc{Dom}} & pcm{\_}RMSenergy{\_}sma{\_}stddev & 0.447 \\ \bottomrule
\end{tabular} \\
\vspace{2.5mm}
\caption{Feature-wise \textsc{CCC} rates for top-performing handcrafted FRs. Abbreviation: Dim. - dimensionality, Val - valence, Aro - arousal, Dom - dominance, CCC - Concordance Correlation Coefficient.}
\label{tab:FR_CC_top}
\end{table}

\section{Summary and conclusions}
\label{sec:Concl}
This paper focused on sensing students' emotion in a remote learning environment. This is a highly challenging problem, as unlike face-to-face classroom learning, learning happens asynchronously. Furthermore, it is not trivial to employ separate scenarios like acted emotion, Wizard-of-Oz setup, talk shows or podcasts as traditionally done in the paralinguistic speech processing community for carrying out emotion prediction research, as we were interested in the emotion that is present in the remote learning process. So, in this work, we investigated a setup where spontaneous monologue speech is acquired with in the learning loop within the self-control tasks for aiding remote learning. 

Given those recordings, we first developed the 56 speakers SPOT-ED dataset by segmenting the data into semantically completed chunks; carrying out a sentiment analysis on transcriptions of students speech; and selecting chunks not too short or not too long. We then investigated answers to two fundamental research questions:
\begin{enumerate}
\item whether that spontaneous monologue speech exhibits perceptible variation in valence, arousal, and dominance dimensions? To ascertain that, we recruited six human listeners; trained them with an A/B test on the VAM corpus; and annotated SPOT-ED. We found moderately high inter-annotator agreement for all dimensions ($\ge 0.6$) and low average assessment quality. We obtained an inter-annotator agreement that was comparable to that of previous studies conducted on the \textsc{VAM} database.
We also observed that EWE labels obtained by combining the responses of all annotators have a wide range of variation along the emotion dimensions. Taken together, this shows that the spontaneous monologue speech collected as open responses in self-control tasks in the remote learning environment exhibit perceptible variations along the emotion dimensions valence, arousal, and dominance.

\item whether dimensional emotion information can be automatically predicted from that spontaneous monologue speech in a reliable manner? For that, we investigated knowledge-based feature representations and state-of-the-art SSL-based pretrained neural representations. For standalone features, regression studies yielded moderate Spearman's and Pearson's correlations and low root mean square errors. Feature level combination yielded the best performances with moderate-to-high Spearman's and Pearson's correlations and reduced root mean square errors. This shows that dimensional emotion information can be reliably predicted from the spontaneous monologue speech collected as part of self-control tasks. Feature ranking analysis and comparison to the VAM corpus studies revealed similar top ranking features, suggesting there may be common acoustic variations although the speech types are different. This is open for further investigation.
\end{enumerate}
Together these studies indicate that speech-based self-control tasks can be a means to sense student emotion in remote learning environments. Nevertheless, key interdisciplinary research challenges remain, including (a) elucidating how such emotional fluctuations should be interpreted within remote learning contexts, and (b) determining how this information can be used — either independently or in conjunction with other data sources within the learning environment — to inform instructional design and to generate meaningful, actionable feedback for both teachers and learners.

\section{Acknowledgments}
This work in its entirety was carried out at FFHS, Brig and Idiap. At Idiap, this work was partially funded by the SNSF through the Bridge Discovery project EMIL (grant no. $40$B$2-0{\_}194794$) and 
the Innosuisse grant agreement no. PFFS$-21-47$ (IICT flagship).
We would like to thank the group of annotators who were involved in dimensional emotion labeling.

\bibliographystyle{IEEEtran}
\bibliography{ref,ICASSP_2024}


\vfill

\end{document}